\documentclass[aps,prd,
twocolumn
nofootinbib,
superscriptaddress,
floatfix
]{revtex4-2}
\usepackage{dcolumn}
\usepackage{bm}
\usepackage{indentfirst}
\usepackage{amsmath}
\usepackage{threeparttable}
\usepackage{graphicx}
\usepackage{amsmath}
\usepackage{booktabs}
\usepackage{array}
\usepackage{footnote}
\makesavenoteenv{tabular}
\usepackage{listings}
\usepackage{amssymb}
\usepackage{subfigure}
\usepackage{amssymb}
\usepackage{epstopdf}
\usepackage{adjustbox}
\usepackage{hyperref}
\usepackage[section]{placeins}
\usepackage{graphicx}
\usepackage{array}

\usepackage[utf8]{inputenc}
\hypersetup{
    colorlinks=true,
    linkcolor=red,
    citecolor=blue,
}
\usepackage{color}
\usepackage[T1]{fontenc}
\usepackage{txfonts}
\usepackage{orcidlink}

\begin{document}

\title{Rapid Parameter Estimation for Extreme Mass Ratio Inspirals Using Machine Learning }

\author{Bo Liang}
 \altaffiliation[]{These authors contributed equally to the work.}
 \affiliation{Center for Gravitational Wave Experiment, National Microgravity Laboratory, Institute of Mechanics, Chinese Academy of Sciences, Beijing 100190, China}
 \affiliation{Shanghai Institute of Optics and Fine Mechanics, Chinese Academy of Sciences, Shanghai 201800, China}
 \affiliation{Taiji Laboratory for Gravitational Wave Universe (Beijing/Hangzhou), University of Chinese Academy of Sciences (UCAS), Beijing 100049, China}
\author{Hong Guo}
\altaffiliation[]{These authors contributed equally to the work.}
 \affiliation{Escola de Engenharia de Lorena, Universidade de São Paulo, 12602-810, Lorena, SP, Brazil}

\author{Tianyu Zhao}
 \affiliation{Center for Gravitational Wave Experiment, National Microgravity Laboratory, Institute of Mechanics, Chinese Academy of Sciences, Beijing 100190, China}

\author{He Wang}
 \affiliation{International Centre for Theoretical Physics Asia-Pacific (ICTP-AP), University of Chinese Academy of Sciences (UCAS), Beijing 100049, China}
 \affiliation{Taiji Laboratory for Gravitational Wave Universe (Beijing/Hangzhou), University of Chinese Academy of Sciences (UCAS), Beijing 100049, China}
  
\author{Herik Evangelinelis}
 \affiliation{Escola de Engenharia, Universidade Federal Fluminense, 24210-240, Niterói, RJ, Brazil}

\author{Yuxiang Xu}
 \affiliation{Center for Gravitational Wave Experiment, National Microgravity Laboratory, Institute of Mechanics, Chinese Academy of Sciences, Beijing 100190, China}
 \affiliation{Shanghai Institute of Optics and Fine Mechanics, Chinese Academy of Sciences, Shanghai 201800, China}
 \affiliation{Taiji Laboratory for Gravitational Wave Universe (Beijing/Hangzhou), University of Chinese Academy of Sciences (UCAS), Beijing 100049, China}

\author{Chang liu}
 \affiliation{Center for Gravitational Wave Experiment, National Microgravity Laboratory, Institute of Mechanics, Chinese Academy of Sciences, Beijing 100190, China}
 \affiliation{National Space Science Center, Chinese Academy of Sciences, Beijing 100190, China}

\author{Manjia Liang}
 \affiliation{Center for Gravitational Wave Experiment, National Microgravity Laboratory, Institute of Mechanics, Chinese Academy of Sciences, Beijing 100190, China}

\author{Xiaotong Wei}
 \affiliation{Center for Gravitational Wave Experiment, National Microgravity Laboratory, Institute of Mechanics, Chinese Academy of Sciences, Beijing 100190, China}

\author{Yong Yuan}
 \affiliation{Center for Gravitational Wave Experiment, National Microgravity Laboratory, Institute of Mechanics, Chinese Academy of Sciences, Beijing 100190, China}

\author{Peng Xu}
 \email{xupeng@imech.ac.cn}
 \affiliation{Center for Gravitational Wave Experiment, National Microgravity Laboratory, Institute of Mechanics, Chinese Academy of Sciences, Beijing 100190, China}
 \affiliation{Key Laboratory of Gravitational Wave Precision Measurement of Zhejiang Province, Hangzhou Institute for Advanced Study, UCAS, Hangzhou 310024, China}
 \affiliation{Taiji Laboratory for Gravitational Wave Universe (Beijing/Hangzhou), University of Chinese Academy of Sciences (UCAS), Beijing 100049, China}
 \affiliation{Lanzhou Center of Theoretical Physics, Lanzhou University, Lanzhou 730000, China}
 
\author{Minghui Du}
 \email{duminghui@imech.ac.cn}
 \affiliation{Center for Gravitational Wave Experiment, National Microgravity Laboratory, Institute of Mechanics, Chinese Academy of Sciences, Beijing 100190, China}

\author{Wei-Liang Qian}
 \affiliation{Escola de Engenharia de Lorena, Universidade de São Paulo, 12602-810, Lorena, SP, Brazil}
 
\author{Ziren Luo}
 \affiliation{Center for Gravitational Wave Experiment, National Microgravity Laboratory, Institute of Mechanics, Chinese Academy of Sciences, Beijing 100190, China}
 \affiliation{Key Laboratory of Gravitational Wave Precision Measurement of Zhejiang Province, Hangzhou Institute for Advanced Study, UCAS, Hangzhou 310024, China}
 \affiliation{Taiji Laboratory for Gravitational Wave Universe (Beijing/Hangzhou), University of Chinese Academy of Sciences (UCAS), Beijing 100049, China}

\begin{abstract}
Extreme-mass-ratio inspiral (EMRI) signals pose significant challenges in gravitational wave (GW) astronomy owing to their low-frequency nature and highly complex waveforms, which occupy a high-dimensional parameter space with numerous variables.
Given their extended inspiral timescales and low signal-to-noise ratios, EMRI signals warrant prolonged observation periods.
Parameter estimation becomes particularly challenging due to non-local parameter degeneracies, arising from multiple local maxima, as well as flat regions and ridges inherent in the likelihood function.
These factors lead to exceptionally high time complexity for parameter analysis while employing traditional matched filtering and random sampling methods.
To address these challenges, the present study applies machine learning to Bayesian posterior estimation of EMRI signals, leveraging the recently developed flow matching technique based on ODE neural networks.
Our approach demonstrates computational efficiency several orders of magnitude faster than the traditional Markov Chain Monte Carlo (MCMC) methods, while preserving the unbiasedness of parameter estimation.
We show that machine learning technology has the potential to efficiently handle the vast parameter space, involving up to seventeen parameters, associated with EMRI signals.
Furthermore, to our knowledge, this is the first instance of applying machine learning, specifically the Continuous Normalizing Flows (CNFs), to EMRI signal analysis.
Our findings highlight the promising potential of machine learning in EMRI waveform analysis, offering new perspectives for the advancement of space-based GW detection and GW astronomy.

\end{abstract}

\maketitle


\section{Introduction}\label{sec=intro}

Gravitational wave (GW) physics and astronomy, emerging in recent decades as a transformative branch of astronomy and cosmology, promise to revolutionize our understanding of the universe. The advent of GW observations marked the beginning of multimessenger astronomy, enabling concurrent insights from gravitational and electromagnetic signals.  In 2015, the LIGO Scientific Collaboration marked a milestone with the first detection of GWs originating from the merger of two black holes~\cite{LIGOScientific:2014pky}. Since this discovery, ground-based GW observatories, including LIGO and VIRGO, have cataloged nearly one hundred GW events from various binary systems, such as binary black holes, binary neutron stars, and neutron star-black hole pairs~\cite{LIGOScientific:2016aoc, LIGOScientific:2016sjg, LIGOScientific:2018mvr, LIGOScientific:2020ibl, VIRGO:2014yos, KAGRA:2021vkt}. The sensitive frequency range of these ground-based detectors, constrained by seismic and shot noise, spans from 10 Hz to several kilohertz~\cite{KAGRA:2021vkt, Kaiser:2020tlg}. Consequently, they are capable of observing binary systems with a total mass of up to several hundred solar masses.

Space-based GW detectors like the commissioned LISA project~\cite{LISA:2017pwj,Danzmann:1997hm}, as well as future missions such as DECIGO~\cite{Kawamura:2006up}, Taiji~\cite{Hu:2017mde,Gong:2021gvw}, and TianQin~\cite{Gong:2021gvw,TianQin:2015yph}, are engineered to detect GW signals in the millihertz to decihertz bands. These bands contain a diverse array of sources that are often challenging to observe but are rich in scientific potential. Such sources typically involve strong-field dynamics and complex interactions within extreme environments, potentially carrying vast amounts of untapped information. These systems include double white dwarfs~\cite{Korol:2017wah,Kang:2021bmp,Ren:2023yec,Huang:2020rjf}, coalescing massive black hole binaries~\cite{Klein:2015hvg,Ruan:2019tje,Ren:2023yec,Wang:2019ryf}, extreme mass-ratio inspirals (EMRIs)~\cite{Babak:2017tow,Ren:2023yec,Fan:2020zhy}, stochastic GW backgrounds~\cite{Chen:2018rzo,Chen:2023zkb,Ren:2023yec,Liang:2021bde}, and other unmodeled sources. Among these, EMRIs are particularly significant as primary detection targets for these observatories~\cite{Berry:2019wgg}. In an EMRI system, a stellar-mass compact object (the secondary) is captured by and gradually spirals into a central supermassive black hole (the primary) over more than $10^5$ orbits, emitting low-frequency gravitational radiation in the millihertz band. Due to the cumulative effects of thousands to millions of orbital cycles and the prolonged orbital evolution in strong-field and extreme environments that last several months to years, EMRIs provide an exceptional platform for probing strong gravitational fields and near-horizon physics, as well as for precise measurements of theoretical parameters~\cite{Laghi:2021pqk,Gair:2012nm}.

In the field of EMRI parameter estimation, several key challenges significantly hinder the effectiveness of traditional methods like MCMC. Firstly, the long-term nature of EMRI signals requires extremely precise and computationally intensive waveform modeling, demanding sustained analytical efforts over periods that can extend for months or years~\cite{Barack:2018yvs,Pound:2021qin}. 
Secondly, the high-dimensional parameter space inherent to EMRIs, typically consisting of more than a dozen interrelated variables, adds considerable complexity to the analysis~\cite{Gair:2004iv,Chua:2020stf,Katz:2021yft}.
Lastly, analysis of EMRI signals is often plagued by significant {\it degeneracy}, making it challenging to distinguish between different parameters of a given model~\cite{MockLISADataChallengeTaskForce:2009wir,Chua:2021aah,Babak:2009ua}, as well as {\it confusion}, which arises from the difficulty in discriminating between different GW models~\cite{Glampedakis:2005cf,Chua:2021aah,Qiao:2024gfb}.
Both factors lead to significant challenges in identifying the model as well as its underlying physical parameters.
Specifically, the non-local parameter degeneracies can be largely attributed to multiple local maxima, as well as flat regions and ridges inherent in the high-dimensional likelihood function. 
Traditional methods, particularly Markov Chain Monte Carlo (MCMC)~\cite{Gair:2008zc,Babak:2009ua,Ali:2012zz}, struggle with these challenges due to their inherent limitations. 
MCMC is not only computationally expensive, typically requiring the generation of approximately $\sim 10^{40}$ waveform templates~\cite{Gair:2004iv}, but also inefficient in exploring vast parameter spaces.
In practice, only two to four parameters can be accurately estimated, and this often requires setting the initial estimates near the true values~\cite{EMRI_PE, burke2024accuracyrequirementsassessingimportance, saltas2023emri_mc}.
This limitation undermines the ability to thoroughly explore the parameter landscape, frequently missing critical global maxima necessary for accurate parameter estimation.
In response, several new tools are under development~\cite{Chua:2021aah,Chua:2022ssg,saltas2023emri_mc,Burke:2023lno}.
However, at present, most methods are still incapable of meeting the stringent demands of EMRI parameter estimation, highlighting the need for more efficient and robust deep learning techniques.

Deep learning has rapidly become an essential tool in scientific discovery, yielding significant advancements across a wide range of fields, from physics to biology~\cite{bambi_advances_2021,zhao_dawning_2023}. 
In the realm of GW astronomy, deep learning has proven particularly effective in data analysis, streamlining both the detection and parameter estimation of ground-based binary black hole (BBH) events~\cite{Huerta:2019rtg,Gabbard:2019rde,Gabbard:2017lja,Xia:2020vem,Dax2023FlowMF, Dax_2021, Dax_2023}. 
Its application extends to space-based observations, where it has been instrumental in rapidly identifying and characterizing massive black hole binary (MBHB) events, contributing to a deeper understanding of these cosmic phenomena~\cite{ruan_rapid_2023,Du2023AdvancingSG, liang2024rapidparameterestimationmerging}. 
Additionally, the scope of machine learning techniques has broadened to include extreme mass ratio inspiral (EMRI) data analysis. 
Current research has made strides in denoising~\cite{Zhao:2022qob}, detecting~\cite{zhang_detecting_2022,Zhao:2023ncy}, and performing point estimation of EMRI parameters~\cite{Yun:2023aqa}, which are critical in interpreting these complex signals. 
However, despite these advancements, a significant challenge remains in the Bayesian posterior estimation of EMRI signals.
This difficulty stems largely from the inherent complexity of EMRI data, characterized by protracted signal durations and intricate waveform structures, which continue to test the limits of existing machine learning methodologies.
Overcoming this challenge requires a concerted effort to develop more sophisticated algorithms that can harness the full potential of deep learning for comprehensive EMRI analysis, pushing the boundaries of what is currently achievable in GW astronomy.

In this study, we develop a deep learning approach specifically tailored for the posterior estimation of EMRI signals by employing the recently developed flow matching technique based on ODE neural networks.
Our method not only enhances computational efficiency, but also overcomes the issue of MCMC methods getting easily trapped in local optima during parameter estimation. 
Such an approach exhibits computational efficiency that surpasses traditional MCMC methods by several orders of magnitude, all while maintaining unbiased parameter estimation.
It is demonstrated that the machine learning technology employed is capable of effectively managing the extensive parameter space, comprising as many as eleven variables, inherent to EMRI signals.
Moreover, as the method can be adapted to a broader prior range to achieve unbiased parameter estimates, it significantly narrows the parameter space of Bayesian inference. 
As a result, our findings potentially provide an innovative solution for dealing with real EMRI signals in the future. 
Specifically, one first narrows down the parameter space of EMRI using the proposed machine learning procedure, and then samples the parameter space with MCMC methods using the posterior from the machine learning model as a prior.
Such an approach combines the efficiency of machine learning with the precision of traditional Bayesian inference, offering an innovative solution for the analysis of EMRI signals.

\begin{figure}[ht!]
  \centering
  \includegraphics[width=1.0\textwidth]{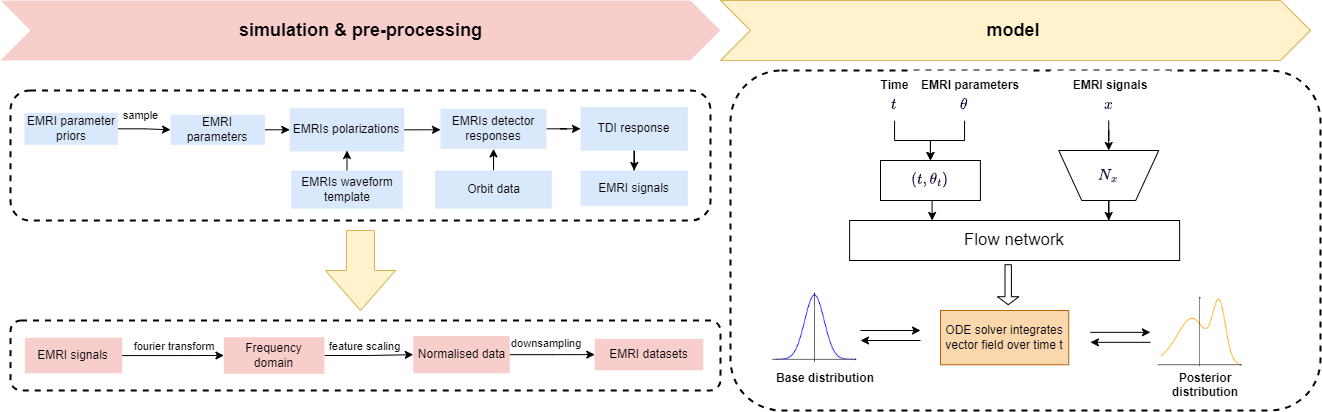}
  \caption{Real-time GW Inference for EMRI Research: An Overview of the Framework Diagram. The left side of the diagram shows the EMRI signal dataset generated based on simulations, detailing the steps of data generation and preprocessing. The right side of the diagram displays the workflow of the machine learning model methods we employ. In sections~\ref{framework}, we provide a detailed exposition of this framework.}
  \label{fig:model}
\end{figure}

The remainder of the paper is organized as follows.
In Sec.~\ref{framework}, we discuss the data generation and preprocessing, laying the foundation for the training and validation of the model. 
Subsequently, we elaborate on the machine learning algorithms specifically tailored for the Bayesian posterior estimation of EMRIs.
The numerical results are presented in Sec.~\ref{result}, where our results are compared against the standard MCMC approach.
Concluding remarks and future perspectives are given in the last section.

\section{Methodological Framework}\label{framework}

\subsection{Data Generation and Preprocessing}
In this section, we give a detailed account of the dataset preparation for model training and validation, which consists of the generation of EMRI waveforms, projection to detector responses, and a pre-processing step. 

A prominent challenge in the analysis of EMRI signal lies in the accurate and high-fidelity modeling of waveform template, which relies on the theories of gravitational self-force and black hole perturbation, \emph{etc}~\cite{Barack:2018yvs,Pound:2021qin}.
To achieve the sub-radian phase accuracy required by the scientific explorations via EMRIs~\cite{Hinderer:2008dm}, perturbative expansions of the binary metric have already been applied at the first order for general orbits around Kerr black holes~\cite{vandeMeent:2017bcc}, with significant progress also made in second-order calculations~\cite{Pound:2019lzj,Warburton:2021kwk}. 
However, the efficiency of numerical computations for gravitational self-force is insufficient to meet the demands of data analysis, hence the rapid generation of approximate ``kludge'' models have been developed~\cite{Barack:2003fp,Babak:2006uv,Chua:2015mua,Chua:2017ujo}.
Early analytic kludge (AK) models~\cite{Barack:2003fp}, while sacrificing some accuracy, offer significantly improved computational efficiency. 
Numerical kludge (NK) models~\cite{Gair:2005ih,Babak:2006uv}, on the other hand, achieve high-fidelity EMRI waveforms by fitting perturbative calculations.
Augmented analytic kludge (AAK) models~\cite{Chua:2015mua,Chua:2017ujo,Chua:2020stf} incorporate the NK model, enabling the generation of high-precision EMRI waveforms without significantly increasing computational costs. 
To date, the most advanced computational framework named \texttt{FastEMRIWaveforms} (\texttt{FEW})~\cite{Katz:2021yft,Chua:2020stf} can rapidly compute fully relativistic EMRI waveforms in the time domain.  As the first step of data preparation, we employ the AAK model implemented in \texttt{FEW} to generate the polarizations of EMRI waveforms.

For the second step, we adopt the generic time-domain response function \texttt{FastLISAResponse}~\cite{Katz_2022} to project the polarizations into the detector responses of LISA, and the resulting EMRI signals are output in the form of second-generation Time Delay Interferometry (TDI) variables~\cite{Martens_2021}. 
Each sample in the dataset includes the  TDI-$\{A, E\}$ variables with a duration of 2 years, calculated at a sampling frequency of 0.1 Hz. 
We employ the Graphics Processing Units (GPUs) to accelerate the computation of waveforms and LISA response functions~\cite{Katz_2022}, achieving the generation of each EMRI signal in less than one second~\cite{burke2024accuracyrequirementsassessingimportance}.
The whole dataset consists of 500,000 samples, with the source parameters randomly drawn from the priors specified by Table~\ref{table:priors}.
In our research, we have come to deeply appreciate the importance of the effectiveness, reliability, and scalability of machine learning algorithms in achieving long-term research objectives. 
To ensure the realization of these goals, we have adopted a robust, step-by-step approach, continuously improving and optimizing based on existing research. 
In this process, we have carefully adjusted and compromised our prior assumptions. In particular, to enhance the accuracy of posterior estimation, we have reasonably restricted the prior range of some intrinsic parameters of EMRI signals.
In subsection~\ref{result}, we have conducted a detailed discussion.

\begin{table*}[h]
    \footnotesize
        \begin{threeparttable}
    \tabcolsep 5pt 
    \caption{\textbf{The prior distributions of EMRI source parameters used in this work. 
    For each parameter the table presents  its lower and upper bounds, and a uniform distribution is assumed between these bounds.
}}
    \label{table:priors}
    \begin{tabular*}{\textwidth}{clccc}
    \toprule
        \hline
        \textbf{Parameter} & \textbf{Description} & \textbf{Prior Lower Bound} & \textbf{Prior Upper Bound} \\
        \hline
        $\mathcal{M} [M_\odot]$ &Primary mass &  $9 \times 10^5 M_\odot$ & $1.1 \times 10^6 M_\odot$ \\
        $\mu$ &Secondary mass & $9 M_\odot$ & $11  M_\odot$ \\
        $a$ &Primary spin &  0.80 & 0.99 \\
        $p_0$ &Semi-latus rectum &  9.10 & 9.30 \\
        $e_0$ &Orbital eccentricity &  0.1 & 0.3 \\
        $Y_0$ &Orbital inclination parameter&  0.59 & 0.79 \\
        $\theta_S$ [{\rm rad}]&Polar angle of source orientation&  0 rad & $\pi$ rad \\
        $\phi_S$ [{\rm rad}]&Azimuthal angle of source orientation&   0 rad & $2\pi$ rad \\
        $\theta_K$ [{\rm rad}]&Polar angle of orbital angular direction &   0 rad & $\pi$ rad \\
        $\phi_K$ [{\rm rad}]& Azimuthal angle of orbital angular direction & 0 rad  & $2\pi$ rad \\
        \hline
        \bottomrule
    \end{tabular*}
    \end{threeparttable}
\end{table*}

To reasonably reduce the dimensionality of data, we further incorporate a pre-processing step. 
Frequency domain analysis~\cite{bayle2022overview,speri2022roadmap,burke2024accuracyrequirementsassessingimportance} proves to be an exceptionally effective method for estimating parameters in EMRIs. 
Therefore, the time-domain samples are converted to the frequency domain using the Fast Fourier Transform (FFT) algorithm. 
The data of EMRIs are inherently challenging to process due to their vast sizes~\cite{Babak_2015}, 
and this issue is even more pronounced for machine learning, for which the models are trained on batched data. 
 After FFT, the length of each sample is up to millions, necessitating methods such as downsampling or pooling. 
To reduce the complexity of the data while still retaining the key information, we adopted a max-pooling strategy~\cite{paszke2019pytorchimperativestylehighperformance}. 
By setting a pooling window of 512 and a stride, we effectively reduced the length of each signal to 8,196 points. 
Prior to training, the 500,000 samples are stored on the hard disk, establishing a robust data foundation for later calculations. 

\subsection{Machine Learning Framework}

We begin this section by reviewing our methodology, focusing on the use of CNFs for fast and accurate EMRIs parameter estimation. 
Currently, many successful attempts aimed at enhancing parameter estimation through machine learning have adopted the 
neural posterior estimation (NPE) approach~\cite{papamakarios2018fastepsilonfreeinferencesimulation, Du2023AdvancingSG, liang2024rapidparameterestimationmerging, Dax_2021, gebhard2023inferringatmosphericpropertiesexoplanets} using discrete normalizing flows (DNFs)~\cite{rezende2015variational, JMLR:v22:19-1028}. 
In theory, the application of NPE to EMRIs is feasible.
However, recent research in simulation-based inference (SBI) suggests that Flow Matching (FM)~\cite{lipman2022flow} with CNFs~\cite{chen2019neuralordinarydifferentialequations} is a more promising approach.
This method is known as Flow Matching Posterior Estimation (FMPE)~\cite{Dax2023FlowMF, lipman2022flow}.
FMPE offers a more rapid training process and superior accuracy compared to NPE~\cite{liang2024rapidparameterestimationmerging, Dax2023FlowMF, gebhard2023inferringatmosphericpropertiesexoplanets}.

\subsubsection{Continuous Normalizing Flows}

CNFs achieve the transformation from a base distribution to a complex target distribution via a vector field \( v_{t,x} \) defined over continuous time. 
This transformation is characterised by an Ordinary Differential Equation (ODE) and is realized by integration. The vector field \( v_{t,x} \) follows this ODE:
\begin{equation}
    \frac{d}{dt} \psi_{t,x}(\theta) = v_{t,x}(\psi_{t,x}(\theta)).
\end{equation}
During the training of CNFs, we initiate from the base distribution and employ an ODE solver to derive the target distribution. 
To ensure consistency with actual data, train the model parameters by log-likelihood.
However, due to the cost of integrating ODEs, the optimization process for maximum likelihood can become exceedingly costly, making it challenging to implement in practice~\cite{Dax2023FlowMF}.

\subsubsection{Flow Matching Posterior Estimation}

The emergence of FM provides a new perspective for training CNFs. Dax et al.~\cite{Dax2023FlowMF} have applied the FM to simulation-based inference, offering a more efficient training objective that helps address the computational challenges inherent in traditional methods.
The loss function of FMPE is given by the following equation:
\begin{equation}
L_{\rm FMPE} = E_{t \sim p(t), \theta_1 \sim p(\theta), x \sim p(x|\theta_1), \theta_t \sim p_t(\theta_t|\theta_1)} \parallel v_{t,x}(\theta_t) - u_t(\theta_t|\theta_1) \parallel^2. \ \ 
\end{equation}
In the FMPE approach, the vector field \( v_{t,x}(\theta_t) \) is responsible for generates the path to the target probability distribution, while \( u_t(\theta_t|\theta_1) \) adjusts this path based on the conditions of the sample. By defining a Gaussian path with time-dependent parameters, FMPE ensures that at \( t=0 \) and \( t=1 \), the probability path closely approximates the base and target distributions.

\subsubsection{Application of Continuous Normalizing Flows}

In order to accurately estimate parameters for EMRIs, we have conducted customized training on our model. Our model is based on CNF. It is optimized using FMPE technology.
The model framework diagram is shown in Figure~\ref{fig:model}.
The model has two parts: 
(1) The first part is a linear compression network \( N_{x} \), designed to effectively capture the intrinsic structure of the signal by compressing the data \( x \) into a 1024-dimensional feature space. This process not only simplifies the data processing workflow but also enhances the model's ability to capture signal features.
(2) The second part of the model is a flow network composed of 56 residual blocks with decreasing sizes, ranging from 4096 dimensions down to 10 dimensions. It receives the embedded tuple \( (x, t, \theta_t) \) and predicts the vector field \( v_{t,x}(\theta_t) \).
We trained the model for 100 epochs, with a batch size of 1024 samples per epoch. The initial learning rate was set to 0.00005, and we employed a cosine annealing strategy to gradually reduce the learning rate to zero as training progressed. The training was conducted on an NVIDIA A800 GPU, the entire training process took approximately 1 hour.
Once our model is trained, the inference process for a single EMRI signal can be completed in just tens of seconds.

\section{Numerical Results}\label{result}

\subsection{Bayesian inference}\label{BayesianResult}

Bayesian statistical inference along with MCMC sampling have been widely adopted in GW source detection and parameter estimation, and is nowadays one of the benchmark approaches in this area, even for the global analysis of multiple sources~\cite{Littenberg:2023xpl,katz2024efficientgpuacceleratedmultisourceglobal}. 
Since FMPE is designed to simulate the Bayesian posterior distribution of parameters, its performance could be more clearly demonstrated if the Bayesian results for the same event are presented for comparison.
The framework of Bayesian statistical inference is built on the Bayes' theorem, which, neglecting the normalization factor~\cite{burke2024accuracyrequirementsassessingimportance}, can be expressed in logarithmic form as:
\begin{equation}
\log p(\theta|d) = \log p(d|\theta) + \log p(\theta).
\end{equation}
The first term on the right hand side \( \log p(d|\theta) \) represents the likelihood function.  The likelihood function  describes the probability of observing the data stream \( d \) given the parameters \( \theta \), which takes a Whittle form under the assumption of Gaussian and stationary noise~\cite{Whittle1957CurveAP}.  The second term \( \log p(\theta) \) is the prior probability distribution of the parameters \( \theta \) given in Table~\ref{table:priors}.

In the parameter estimation of EMRI signals, we face challenges caused by non-local parameter degeneracies, which lead to multiple local maxima on the likelihood surface. These local maxima make the EMRI signal extremely sensitive to small variations in parameters, thus accurately identifying the global maximum in the likelihood function becomes particularly difficult. Currently, when estimating EMRI parameters based on the MCMC method~\cite{Ali_2012,Karnesis_2023,burke2024accuracyrequirementsassessingimportance}, it is usually necessary to search within a small vicinity of the true parameter values, or to infer only a few parameters of the EMRI.

To highlight the potential and advantages of machine learning methods in EMRI parameter estimation, we implement the conventional Bayesian method via an advanced MCMC sampler \texttt{Eryn}~\cite{Karnesis_2023}  based on the \texttt{Emcee}~\cite{Foreman_Mackey_2013} code  to sample from the posterior distribution, which is one of the widely used tools in the current space-base GW data analysis. 
We performed Bayesian inference on the injected EMRI signals to test the performances of both MCMC and our machine learning methods.
We made two settings for MCMC tests, only differing in the choices of starting points~\cite{EMRI_PE}: 
\begin{enumerate}
    \item MCMC (1):  The starting points were selected within a very small neighborhood of the true parameter values (within a range of \( \times 10^{-7}\) times the true values);
    \item MCMC (2): The starting points were randomly sampled according to the priors.
\end{enumerate}

\begin{figure}[h]
  \centering
  \includegraphics[width=1.0\textwidth]{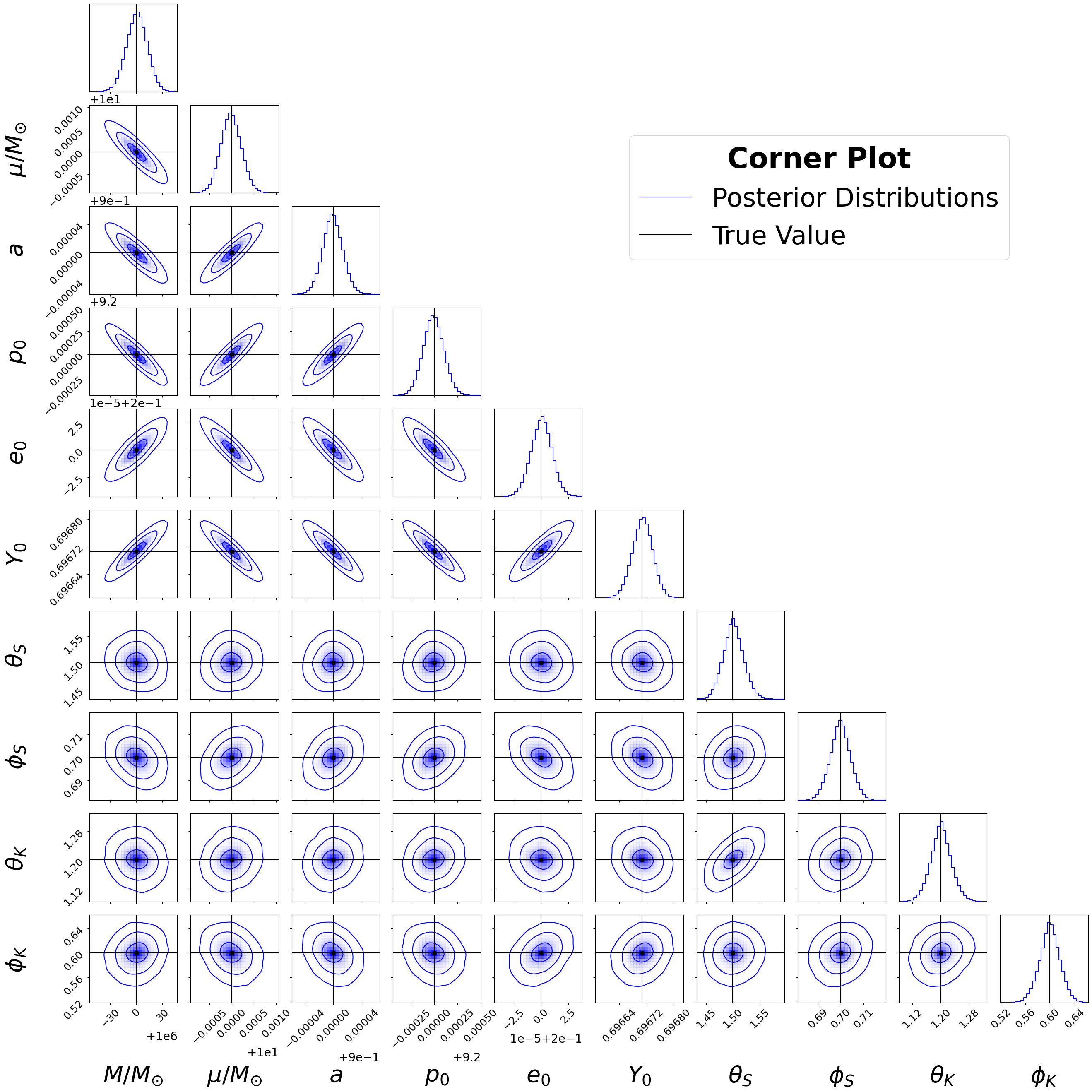}
  \caption{The figure displays the posterior distribution for an EMRI, which was obtained by setting the starting points of the MCMC method close to the true parameter values.
  The analyzed EMRI signal has a signal-to-noise ratio (SNR) of approximately 67 and is based on two years of observational data. In the figure, the blue line represents the posterior probability distribution of the EMRI parameters, while the black line indicates the true parameter values.}
  \label{fig:corner_mcmc}
\end{figure}

By running MCMC according to the setting (1), we obtain the posterior distributions of parameters  shown in  Figure~\ref{fig:corner_mcmc}, where the contours from inside to outside represent the 1 $\sigma$, 2 $\sigma$, and 3 $\sigma$ confidence intervals, respectively, and the black lines indicate the true parameters underlying the injected signal.
Unsurprisingly, all the median values of parameters are perfectly coincident with the truths, demonstrating that when the starting points are close to the true values, MCMC can efficiently and correctly identify the injected signals.
While, given that in practical data analysis  we do not have prior access to the true parameters, we explored the capability of MCMC based on the  setting (2), which involves randomly sampling starting points from the prior distribution. 
The results of the second setup are shown in Figure~\ref{fig:mcmc}. 
For this case, the MCMC method failed to recover the true parameters, with most of  the ``posteriors'' clustered near the boundaries set by the priors. 
Figure~\ref{fig:mcmc} also shows the convergence plot of the MCMC chains, indicating that the MCMC ultimately only converged to a local optimum and failed to find the global optimum~\cite{Gair:2004iv, Babak:2009ua, Babak_2015}.
Both settings of the MCMC method utilized GPU acceleration to enhance computational speed, and they still required approximately two days to complete the posterior estimation of EMRI signals.
Although this issue may be resolved by, for example,  incorporating the parallel tempering mechanism~\cite{PhysRevLett.57.2607} to escape from local maxima,  while this would imply  at least a several-fold increase in the  computation  time, making its practical application still less preferable. 

\subsection{The present approach}~\label{result-f}

\begin{figure}[htb]
  \setlength{\arrayrulewidth}{0.2pt} 
  \centering
  \includegraphics[width=0.4\textwidth]{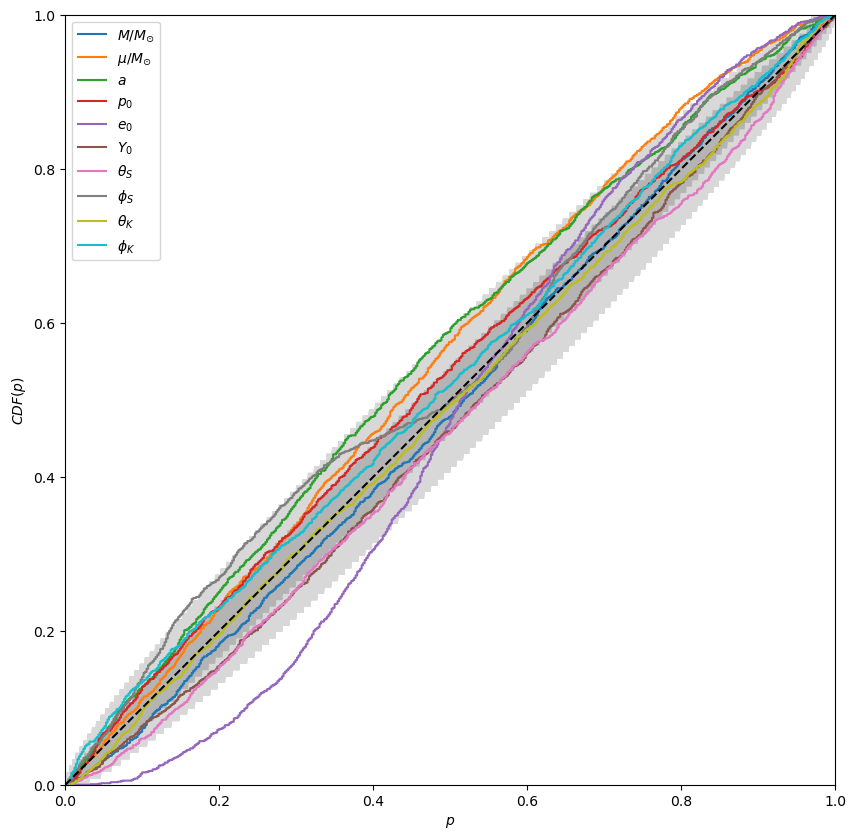}
  \caption{P-P plot for a set of 1000 injections, as analyzed by our machine learning model. We calculate the true percentile from the marginalised posterior and plot the CDF.}
  \label{fig:pp}
\end{figure}

To evaluate our machine learning model, we first tested it on 1,000 EMRI signals randomly generated accroding to the priors. 
By drawing posterior samples of these signals using our model, we constructed a P-P plot shown in  Figure~\ref{fig:pp}. 
In principle,  the cumulative distribution function (CDF) of the posterior percentile values for each parameter should approximate the diagonal line, and is just the case of our  results,
thereby validating the unbiasedness of our estimations over the whole parameter range in consideration.

\begin{figure}[htb]
  \centering
  \includegraphics[width=1.0\textwidth]{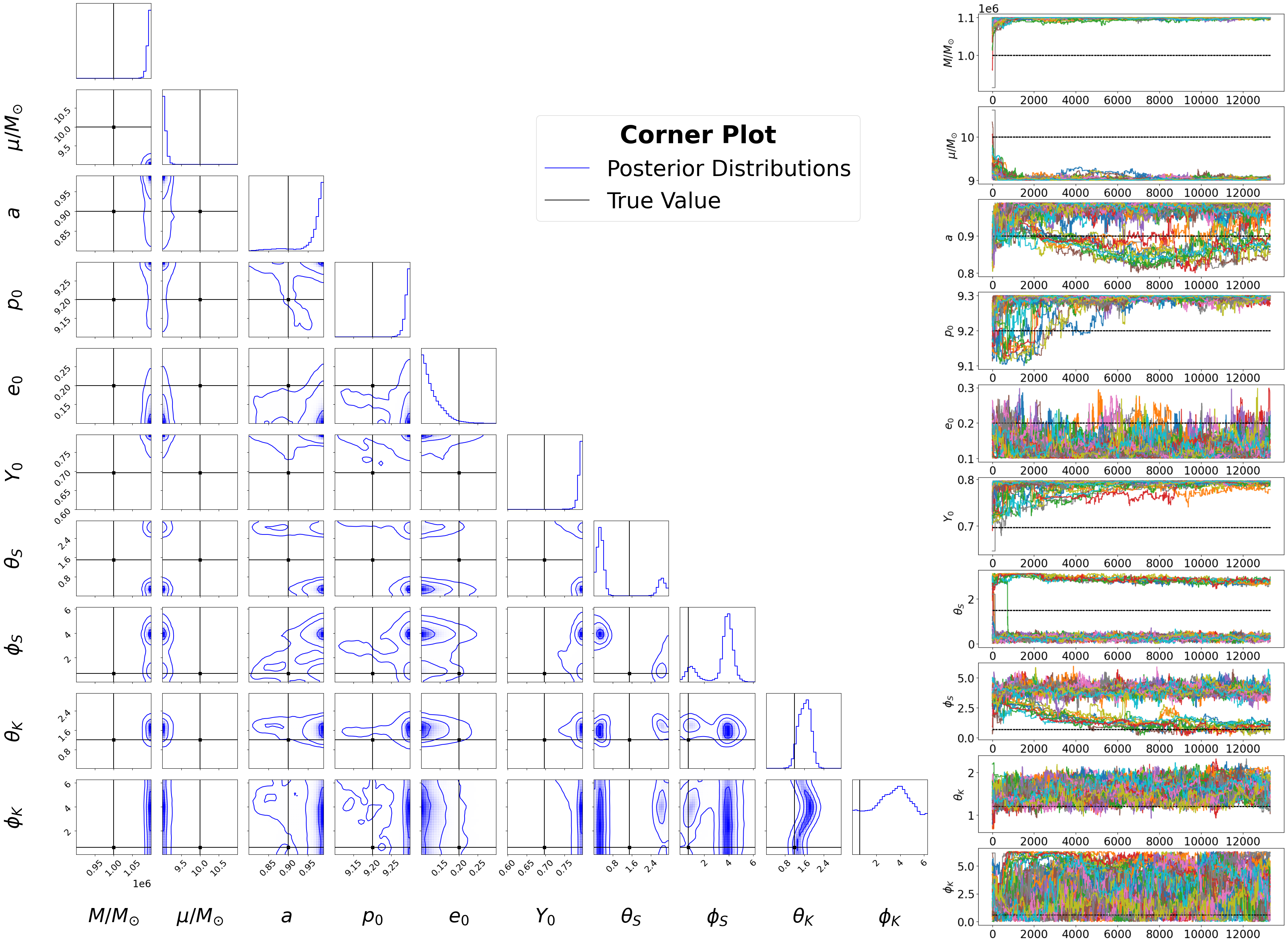}
  \caption{
  The left side of the figure displays for an EMRI. The starting points for the MCMC were obtained by random sampling from the prior distribution.
  The analyzed EMRI signal has a SNR of approximately 67 and is based on two years of observational data. In the figure, the blue line represents the posterior probability distribution of the EMRI parameters, while the black line indicates the true parameter values. The plot in the right shows the MCMC chains converging.
  The right side of the figure displays the convergence of the MCMC chain.
  }
  \label{fig:mcmc}
\end{figure}

\begin{figure}[htb]
  \centering
  \includegraphics[width=1.0\textwidth]{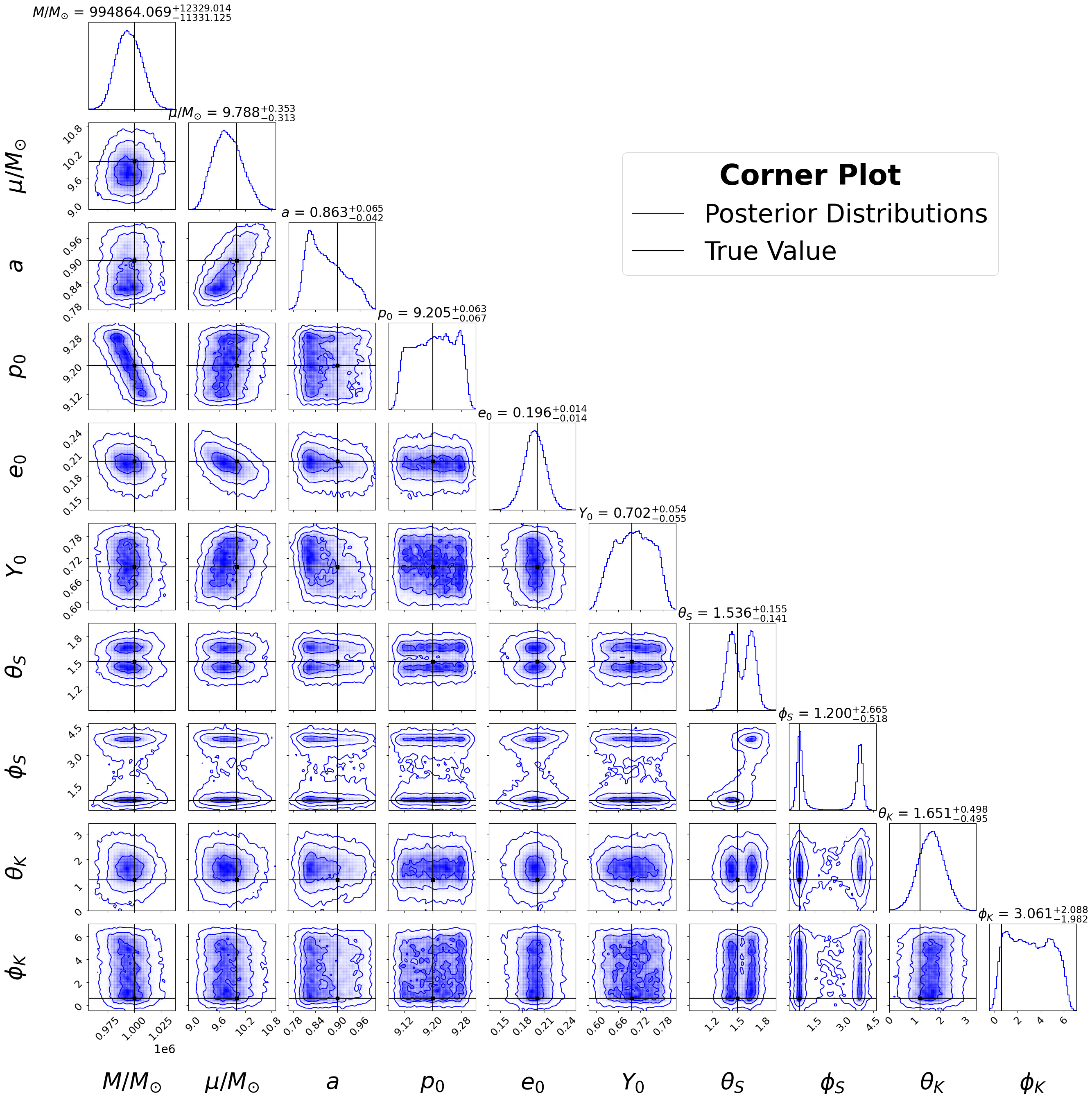}
  \caption{The figure displays the posterior distribution for an EMRI. The analyzed EMRI signal has a SNR of approximately 67 and is based on two years of observational data. In the figure, the blue line represents the posterior probability distribution of the EMRI parameters, while the black line indicates the true parameter values.}
  \label{fig:corner_ml}
\end{figure}

In comparison to the MCMC method, we  also performed  parameter estimation on  the same  EMRI signal as Sec.~\ref{result}. 
Since our model can not exploit the advantage of knowing the true parameters  priorly,  its application scenario is essentially identical to MCMC (2) (\emph{i.e.} the  more realistic one).
 As is shown in Figure~\ref{fig:corner_ml}, 
 our model successfully reconstructed the posterior distribution of the signal, with all the true parameters lying within the 1 $\sigma$ (or at most 2 $\sigma$) ranges. 
 This is in stark contrast to the result of MCMC (2) (Figure~\ref{fig:mcmc}), where  the estimations are severely biased. 
This comparison provided us with an intuitive overview on the performance of both methods. 
FMPE exhibits a strong capability to escape from local maxima, and this is achieved in a very short time scale. 
To some extent, this may suggest that FMPE has the potential to become  a truly viable methodology in terms of both efficiency and accuracy in the area of EMRI posterior estimation.

To further assess the reliability of  our model in terms of  reconstructing  posterior distributions, it is helpful to  compare with MCMC (1),  whose posterior might be  regarded  as a benchmark. 
Our machine learning process typically generates posterior distributions with a broader range than those of \texttt{Eryn}, 
however it has effectively narrowed down the parameter ranges compared to the priors, 
especially for the sky position parameters (as can be seen in the first peaks of $\theta_S$ and $\phi_S$), making it possible to perform rapid localization of EMRIs via our model. 
Besides, although the phenomenon of multiple peaks is observed in the parameters $\theta_{s}$ and $\phi_{s}$, which is not the same as the results of \texttt{Eryn} searching near the truth value. 
We believe that this phenomenon is reasonable because \texttt{Eryn} searches near the truth value and may not have explored all potential peaks in the parameter space.

Through the above comparisons, by integrating   the features  of MCMC and machine learning, we can gain insightful ideas to further refine the results of both methods.
Looking ahead, in dealing with real EMRI signals, we can adopt the following strategy: first, use the machine learning process to quickly narrow down the range of the parameter space, and then apply the MCMC method for more refined sampling, using the posterior distribution of the machine learning model as a prior. This approach combines the efficiency and the ability to break local maxima of machine learning with the precision of Bayesian inference, providing an innovative solution for the analysis of EMRI signals.


\begin{table*}[htb]
    \large
    \centering 
    \footnotesize
    \setlength{\arrayrulewidth}{0.4pt} 
    \renewcommand{\arraystretch}{1.5}
        \begin{threeparttable}
    \tabcolsep 15pt 
    \caption{\textbf{The table compares the parameters injected a prior in the real case with those recovered by FMPE and emcee. The recovered values are accompanied by their 1 $\sigma$ confidence regions.
}}
    \label{table:result}
    \begin{tabular*}{\textwidth}{clccc}
    \toprule
        \hline
        \textbf{Parameter} & \textbf{Description} & \textbf{Injected value} & \textbf{FMPE} & \textbf{Eryn} \\
        \hline
        $\log_{10}{\mathcal{M}} [M_\odot]$ & Primary mass & 6 &  $5.998_{-0.005}^{+0.005}$ & $6.041_{-0.001}^{+0.000}$ \\
        $\mu [M_\odot]$ & Secondary mass &  10 &  $9.788_{-0.313}^{+0.353}$ & $9.017_{-0.013}^{+0.031}$\\
        $a$ & Primary spin & 0.9 &  $0.863_{-0.042}^{+0.065}$ & $0.979_{-0.028}^{+0.008}$\\
        $p_0$ & Semi-latus rectum & 9.2 &  $9.205_{-0.067}^{+0.006}$ & $9.296_{-0.008}^{+0.003}$ \\
        $e_0$ & Orbital eccentricity & 0.2 &  $0.196_{-0.014}^{+0.014}$ & $0.121_{-0.016}^{+0.032}$ \\
        $Y_0$ & Orbital inclination parameter& 0.7 &  $0.702_{-0.055}^{+0.054}$ & $0.795_{-0.004}^{+0.001}$ \\
        $\theta_S [{\rm rad}] $ & Polar angle of source orientation& 1.5 &  $1.536_{-0.141}^{+0.155}$ & $0.308_{-0.104}^{+2.489}$ \\
        $\phi_S [{\rm rad}] $ & Azimuthal angle of source orientation& 0.7 &  $1.200_{-0.518}^{+2.665}$ & $3.824_{-2.559}^{+0.385}$ \\
        $\theta_K [{\rm rad}] $ & Polar angle of orbital angular direction & 1.2 &  $1.651_{-0.495}^{+0.498}$ & $1.632_{-0.265}^{+0.208}$ \\
        $\phi_K [{\rm rad}] $ & Azimuthal angle of orbital angular direction & 0.6 &  $3.061_{-1.982}^{+2.088}$ & $3.31_{-2.115}^{+1.774}$ \\
        \hline
        \bottomrule
    \end{tabular*}
    \end{threeparttable}
\end{table*}

\section{Concluding Remarks and Perspectives}

The present study applies machine learning algorithms to EMRI posterior estimation, offering a novel approach for Bayesian parameter estimation of EMRI signals.
Our method not only provides a comprehensive statistical framework for analyzing EMRIs but also establishes a new benchmark in space-based GW astronomy.
We have demonstrated the significant potential of machine learning in handling the low-frequency characteristics and vast parameter space of EMRI signals.
Compared to traditional MCMC methods, our machine learning approach efficiently explores a broader parameter space, including regions that are challenging for conventional methods to reach.
In terms of computational speed, our method delivers an order-of-magnitude improvement over MCMC techniques.
This work provides new insights into enhancing existing parameter estimation methods, particularly in addressing complex signals characterized by non-local parameter degeneracies and multi-harmonic overlap.
As machine learning technology continues to advance, we anticipate it will play an increasingly critical role in space-based GW detection, driving further progress in GW astronomy.
Additionally, our findings suggest a promising approach for processing future real EMRI signals.
Specifically, machine learning could be used to narrow the parameter space, followed by MCMC methods, where the prior likelihood distribution is informed by the machine learning-derived posterior.
This strategy effectively combines the efficiency of machine learning with the precision of Bayesian inference, providing a substantially powerful tool for EMRI signal analysis.

\begin{acknowledgments}
This study is supported by the National Key Research and Development Program of China (Grant No. 2021YFC2201901, Grant No. 2021YFC2203004, Grant No. 2020YFC2200100 and Grant No. 2021YFC2201903). 
International Partnership Program of the Chinese Academy of Sciences, Grant No. 025GJHZ2023106GC.
We also gratefully acknowledge the financial support from Brazilian agencies 
Funda\c{c}\~ao de Amparo \`a Pesquisa do Estado de S\~ao Paulo (FAPESP), 
Fundação de Amparo à Pesquisa do Estado do Rio Grande do Sul (FAPERGS),
Funda\c{c}\~ao de Amparo \`a Pesquisa do Estado do Rio de Janeiro (FAPERJ), 
Conselho Nacional de Desenvolvimento Cient\'{\i}fico e Tecnol\'ogico (CNPq), 
and Coordena\c{c}\~ao de Aperfei\c{c}oamento de Pessoal de N\'ivel Superior (CAPES).
\end{acknowledgments}

\bibliography{aipsamp}
\end{document}